\documentclass{elsart}

\usepackage{natbib}

\usepackage{epsfig}

\usepackage{amssymb}
\newcommand{\rsun}{R$_\odot$}

\begin{document}

\begin{frontmatter}



\title{Characterization of the slow wind in the outer corona}
%\thanksref{footnote1}}
%\thanks[footnote1]{This template can be used for all publications in Advances in Space Research.}


\author{Lucia Abbo\corauthref{cor}}
\corauth[cor]{Lucia Abbo}
\ead{abbo@oato.inaf.it}
\author{, Ester Antonucci}
%\thanksref{footnote2}}
\address{INAF-Osservatorio Astronomico di Torino, Via Osservatorio 20, 10025, Pino Torinese, Italy}

%\thanks[footnote2]{Additional information regarding the corresponding author}

%url can be given like this
%\ead[url]{http://authors.elsevier.com/locate/latex}

%\author{Ester Antonucci\thanksref{footnote3}}
%\address{INAF-Osservatorio Astronomico di Torino, Via Osservatorio 20, 10025, Pino Torinese, Italy}
%\thanks[footnote3]{Additional information about the second and third authors}
%\ead{more@email.addresses}

\author{Zoran Miki\'c, Jon A. Linker, Pete Riley and Roberto Lionello}
%\thanksref{footnote4}}
\address{Predictive Science, Inc., 9990 Mesa Rim Road, San Diego, CA 92121, USA}
%\thanks[footnote4]{Additional information about the co-authors}
%\ead{more@email.addresses}

\begin{abstract}
% Text of abstract
The study concerns the streamer belt observed at high
spectral resolution during the minimum of solar cycle 23 with the Ultraviolet Coronagraph Spectrometer (UVCS) onboard SOHO. On the basis of a spectroscopic analysis of the O VI doublet, the solar wind plasma parameters are inferred in the extended corona. The analysis accounts for the coronal magnetic topology, extrapolated through a 3D magneto-hydrodynamic model, in order to define the streamer boundary and to analyse the edges of coronal holes. The results of the analysis allow an accurate identification of the source regions of the slow coronal wind that are confirmed to be along the streamer boundary in the open magnetic field region.

\end{abstract}


\end{frontmatter}

\parindent=0.5 cm

% main text
\section{Introduction}

The origin  of the slow solar wind is still among the open problems
 in solar physics. According to the Ulysses observations,  during
solar minimum the slow solar wind tends to be confined in the heliospheric region corresponding to the extension of the equatorial streamer belt, whereas the fast wind fills most of the heliosphere.
Several authors have studied in the recent minimum solar cycle the physical properties of the streamer belt in corona in 1996-1997 using the remote sensing observations from SOHO (Solar and Heliospheric Observatory), the in-situ observations with  Ulysses
and ground-based instruments
(e.g. Abbo et al. 2002, 2003a, 2003b; Abbo \& Antonucci 2002; Antonucci et al. 1997, 2005, 2006; Dobrzycha et al. 1999; Frazin et al. 2003; Habbal et al. 1997; Marocchi et al. 2001; Noci et al. 1997; Noci \& Gavryuseva 2007; Parenti et al. 2000; Poletto et al. 2002; Raymond et al. 1997; Sheeley et al. 1997; Spadaro et al. 2005; Strachan et al. 2000, 2002; Uzzo et al. 2003;
 Wang et al. 1998;  Zangrilli et al. 1999).\\
Observations from the Ultraviolet
Coronagraph Spectrometer (UVCS; Kohl et al. 1995) onboard SOHO, during solar minimum, have revealed that the oxygen emission shows a strong
depletion in the streamer core combined with a much reduced depletion in the bright regions surrounding the dim core.
This fact has been related in different ways to the origin of the slow wind: if a multiple magnetic structure is present in the inner part
of the large equatorial streamers, the low-speed wind could flow between these sub-streamers and a smaller ion drag due to the low speed in this
 region might explain the core abundance depletion (Noci et al. 1997; Marocchi et al. 2001; Noci \& Gavryuseva 2007); according to an alternative interpretation, slow wind might
  arise from the streamer bright regions identified with streamer legs.
  In the latter case, the OVI core dimming in quiescient streamer has been explained in terms of gravitational settling (Raymond et al. 1997). Ofman (2000) explains the brightening of the streamer-edges as due to an
  enhanced abundance of OVI ions caused by the Coulomb friction with the outflowing protons forming the slow wind.\\
  Theoretical models propose different scenarios for the physical processes of the slow wind. The magneto-hydrodynamic model developed
  by Wiegelmann et al. (2000) indicates that small eruptions at the helmet streamer cusp may incessantly accelerate small amounts
  of plasma without significant changes of the equilibrium configuration of multiple sub-streamers. According to the
  hypothesis put forward by Fisk et al. (1999), the interplay of the differential
  rotation of the photosphere and the super-radial expansion of the magnetic field causes a deposition
  of the magnetic flux at low latitudes that could give origin to reconnection with closed magnetic loops, thereby releasing
  material to form the slow wind. Sheeley et al. (1997) have observed, using data by LASCO onboard SOHO, several small coronal mass ejections in form
   of slow 'streamer blowouts', which have a similar behaviour to the discrete inhomogeneities carried by the slow solar
    wind within the streamer belt.\\
The main issue then is whether the slow wind is coming from open field line regions surrounding the streamer
or there is a substantial contribution from the streamer itself: either the brightest regions of the streamer or the core dimming as observed in OVI.
In order to investigate this problem in depth we need to establish with accuracy the relationship between the intensity
of coronal emission and the magnetic topology, which allows us to distinguish between closed and open field line regions.
An earlier publication by Antonucci et al. (2005) derived the electron density and the outflow velocity
in the coronal regions external to and running along the streamer boundary from a spectroscopic analysis of the
O VI 1032 and 1038 \AA~lines observed with UVCS, by assuming the magnetic
geometry of the flow tube connecting the coronal region to the heliosphere as derived by Wang \& Sheeley (1990).
The present study extends the previous analysis by using a magnetic topology directly inferred from the photospheric magnetic field
 for the dates of the UVCS observations through  the three-dimensional magneto-hydrodynamic (MHD) model  of the global
corona developed by Miki\'c et al. (1999). Moreover, in this work the streamer adjacent regions are studied in more details up to the edge of the coronal hole, pointing out the latitudinal variation of the coronal physical parameters from slow to fast wind.
We derive the HI and OVI kinetic temperature from the data analysis of the spectral lines and the coronal electron density as a function of the outflow velocity through a diagnostic technique described in the next section.

\section{Diagnostic techniques for the coronal plasma}
\label{diagn}
Outflow velocity and electron density of the coronal wind plasma can be deduced from the
emission of
intense ultraviolet spectral lines, such as the O~VI 1032 and 1038 lines. These lines are formed
in the extended corona via collisional
 and radiative excitation
processes. The two components have a
different dependence on the electron density: the collisional process depends
on $n_e^2$, while the radiative process
  depends linearly on electron density $n_e$.
The collisional and radiative components of the O~VI 1032 and 1038 lines in an expanding plasma can be separated by
using the method introduced by Antonucci et al. (2004).
 Since the two lines are emitted by the same ion, the ratios of
the radiative
 and collisional components become independent of the abundance of the element
  and of the ionization
 factor.
The electron
density, averaged along the line-of-sight (l.o.s.), $<n_e>$, is proportional to the ratio of the collisional component, $I_c$,
 to the radiative component, $I_r$, and is a function of the outflow
velocity of the wind, {\bf w}:
\begin{equation}
\label{ne}
<n_e>\,\sim\,\frac{I_c}{I_r}\,<\Phi(\delta\lambda)>,
\end{equation}
where
$<\Phi(\delta\lambda)>=\int_{0}^{\infty}\Psi(\lambda-\lambda_0)\,I_{ex}(\lambda-\delta\lambda,\bf{n})\,d\lambda$
 is the Doppler dimming function averaged along the l.o.s.. This quantity  depends
on the normalized
 coronal absorption profile, $\Psi(\lambda-\lambda_0)$, and on  the intensity of the exciting spectrum, $I_{ex}(\lambda-\delta\lambda,\bf{n})$,
 along the direction of the incident radiation,
$\bf n$. The quantity $\delta\lambda=\frac{\lambda_0}{c}\,{\bf w}\cdot{\bf n}$ is
 the shift of the disk spectrum
introduced by the expansion velocity, {\bf w}, of the coronal absorbing
ions/atoms along the direction $\bf n$ and $\lambda_0$ is the reference wavelength of the
transition. As the wavelength shift increases, the resonantly scattered emission decreases,
 giving origin to the Doppler dimming effect (Beckers \& Chipman 1974; Kohl \& Withbroe 1982; Withbroe et al. 1982;  Noci et al. 1987). By analysing the
 O~VI doublet lines at 1031.93 and
1037.62 \AA, it is
possible to determine electron density and oxygen ion outflow velocities (averaged along the l.o.s. on the plane of sky) up to approximately  500 km s$^{-1}$ including
 the effect of pumping
of the CII lines at 1037.02 and 1036.34 \AA~on the O~VI $\lambda$~1037.61 line (Noci et al. 1987;
 Dodero et al. 1998; Li et al. 1998; Cranmer et al. 1999a; Telloni et al. 2007). The values for the observed disk intensity of the O~VI and C~II lines
  are given by Curdt et al. (2001).\\
There are infinite pairs of values ($n_e,w$) satisfying equation (\ref{ne})
 given the observed line intensities.
Therefore when the plasma is dynamic, in order to measure the coronal electron
 density and the outflow velocity
at the same time, we need a
further physical constraint: the mass flux conservation along the
flow tube connecting the corona to the heliosphere.
That is, we apply the continuity equation:
%\begin{equation}
%\label{masscons}
$n_e \times w \times A\,=\,const$,
%\end{equation}
where $A\,=\,f(r)\,r^2$ is the cross section
 of the flux tube and the quantity $f(r)$, the expansion factor, takes into account the deviation
  from radial expansion. According to this technique, the outflow velocity of oxygen ions  is approximated
  to be that of electron and protons.
  The quantity $n_e \times w$  measured in the heliosphere
by $in-situ$ instruments, can be extrapolated to the corona given a flux
tube geometry. The coronal value of ($n_e,w$) intersects the curves ($n_e \times w, w$)
derived from equation (\ref{ne}), giving physically acceptable solutions for
the electron density
$n_e$ and for the outflow velocity $w$.
Hence, the spectroscopic method described above yields physical results only if the magnetic topology
 of the flux tubes in the corona and
 heliosphere is inferred in
an accurate way.
Antonucci et al. (2005) adopted by approximation
 the magnetic topology derived by Wang and Sheeley (1990) for
the regions adjacent to the streamer, in
the case of the dipolar magnetic field model including an equatorial
current sheet. In the present analysis, we have used global MHD models of the solar corona 
(Miki\'c et al. 1999) to infer the magnetic topology and expansion factors of flux tubes for the specific dates of the UVCS observations (see Section \ref{mhd}). Although it is difficult to quantify the accuracy of these solutions, comparisons between simulated and observed white light images over several solar cycles (e.g. www.predsci.com/corona) suggest that the model reproduces the essential features of the large-scale structure of the solar corona.  

\section{Three dimensional MHD model of the global corona}
\label{mhd}
In order to distinguish between closed and open field regions in the outer corona and derive the magnetic field
 expansion factors to get a detailed description of the flow tube geometry,
necessary for applying the spectroscopic diagnostics described in the previous section,
 the coronal magnetic fields have
been extrapolated from photospheric longitudinal fields on the basis of
the three-dimensional MHD model of Miki\'c et al. (1999).
The code integrates the time-dependent MHD equations in spherical
 coordinates ($r,\theta,\phi$).
The
photospheric magnetic field data (obtained from synoptic magnetic field observations at
 Kitt Peak National Solar Observatory on the days of observation considered in the analysis)
 are used to specify the boundary condition on the
radial component of the magnetic field, $B_r$. The boundary conditions on the velocity are determined
 from the characteristic
equations along $B$.
The plasma temperature and  density
are specified with constraint values at the solar surface. The upper radial
boundary is placed beyond the magnetosonic critical point, typically at $r=30$ R$_{\odot}$~
(Linker et~al. 1999).
The MHD equations are integrated in time until the plasma and magnetic fields settle
 into equilibrium. Time histories of selected global and local parameters are visually
  inspected to ensure that a steady-state solution has been achieved.
Although our most recent model includes a full treatment of the most important energy transport processes in the solar corona (Lionello et al. 2009), in this study, we have employed a simpler polytropic model, in which an adiabatic energy equation with a reduced polytropic index, $\gamma$, is used. The motivation for using a reduced $\gamma$ is the fact that the temperature in the corona does not vary substantially (the limit $\gamma\rightarrow$1 corresponds to an isothermal plasma). A typical choice, used for the polytropic model, is $\gamma$=1.05. We have found that the polytropic solutions closely match the full thermodynamic solutions in terms of the structure and morphology of the coronal magnetic field. Moreover, comparisons of simulated white light images with coronagraph observations taken during total eclipses suggest that the model has reproduced the essential large-scale features of the corona (Mikic et al., 1999).

\section{Observations}

In this study, we have selected three days of
 UVCS observations at mid and low latitudes, performed at high spectral resolution during
solar minimum:
 26, 30, 31 August and 1 September 1996.
All these observations were included in the first Whole Sun Month Campaign and they
 are characterized by a well-defined streamer boundary in
the UVCS field of view (FOV), that is, the UVCS scans cover
 the streamer or part of the streamer and
the transition between
 the streamer
 and the coronal hole.
\begin{table}[!htb]
  \begin{center}
  \caption{\small{Observation date, polar angle (in degrees, counterclockwise from the North Pole),
    altitude range (in solar radii, R$_\odot$) and time interval (U.T.) of observation of the solar minimum coronal streamers observed with UVCS/SOHO.}}
   \vspace{1cm}
    \renewcommand{\arraystretch}{1.2}
    \begin{tabular}[h]{|c|c|c|c|}
      \hline
      Date&  Polar angle   &    Altitude range & Time interval \\
     &  ($\theta^o$)     &  (R$_\odot$)  & (U.T.) \\
      \hline
     26 Aug 1996& 315 & 1.5 -- 3.8 & 18:30:38 -- 03:31:05\\
     30 Aug 1996& 225 & 1.5 -- 3.8 & 16:27:47 -- 02:37:36\\
     31 Aug 1996& 255& 1.5 -- 3.8 & 16:59:29 -- 02:00:00\\
     01 Sep 1996& 285 & 1.5 -- 3.8& 16:50:04 -- 01:50:23\\
      \hline
      \end{tabular}
    \label{tab:obs}
 \end{center}
\end{table}
For each observation, UVCS scanned the coronal region between 1.5 R$_\odot$ to 3.8
R$_\odot$, defined by the polar angle
(in degrees, counterclockwise from the North Pole) of the center of the UVCS FOV and by the
time interval of observation given in Table 1.
The observations  were
obtained  at the following heights:
1.5, 1.6, 1.7, 1.8, 1.9, 2, 2.1, 2.3, 2.5, 2.7, 2.9, 3.3 and 3.8 R$_\odot$.
Each day the observation was performed in approximately 9-10 hours.
We analyse the O~VI doublet lines at 1031.93 and
1037.62 \AA\, detected on the UVCS OVI channel (984--1080 \AA), and the HI Lyman $\alpha$ line at 1216 \AA\, detected on the Ly $\alpha$ channel (1100--1361 \AA).
The slit (the spatial direction of the detector)
is oriented perpendicular to the radial direction defined by the polar angle.
The spectrometer slit was 37 arcmin long, with spatial pixels of 7 arcsec binned in groups of 8 for the O~VI channel and in groups of 4 for the Ly $\alpha$ channel.
 The slit width of the spectrometer,
which determines the spectral
resolution of the observation, was selected to be 50 $\mu$m (corresponding to
14 arcsec and 0.18 \AA) up to 3.3 \rsun~and it was
 300 $\mu$m (corresponding to
84 arcsec and 1.08 \AA) at 3.3 and 3.8 R$_\odot$ for the O~VI channel and it was 50 $\mu$m for the Ly $\alpha$ channel. 
In order to increase the statistics in the analysis, we have grouped together a number of contiguous exposures at
 different distances and the resulting average heliodistances are:
 1.6, 1.85, 2.15, 2.6 and 3.33 R$_\odot$ for streamer and 1.6, 1.85 and 2.15 R$_\odot$ for coronal hole (as shown in Fig. 1).

%\clearpage
\section{Data analysis}

As a first step in the analysis, the variation of the coronal plasma physical parameters
 in open and closed magnetic field line regions is studied.
  It is thus necessary to
 define the streamer boundary that, in this case, takes into account
 the coronal magnetic field derived on the basis of the MHD model. The magnetic topology
 and the expansion factors
as functions of coronal latitudes and heights have been inferred relative to the
  plane of sky for the dates of the UVCS
 observations reported in Table 1. The magnetic field line map is shown in Figure 1 (top panel).
 Black lines
 indicate the heliocentric distances of
 the instantaneous field of view (IFOV) of UVCS (bottom-right panel)
 that covers both the
southern and the northern boundary of the streamer structure, on the 30th of August and 1st of September,
respectively.
We consider only the projection on the plane of the sky because it provides
the most important contribution to the UV line intensities when integrating the emission along the line-of-sight.\\
The photospheric boundary of the
coronal hole (open--field region), inferred from the photospheric magnetic measurements, is indicated as black regions on the disk surface, in the field lines map.  
\begin{figure}
\centerline{\includegraphics[width=.4\textheight]{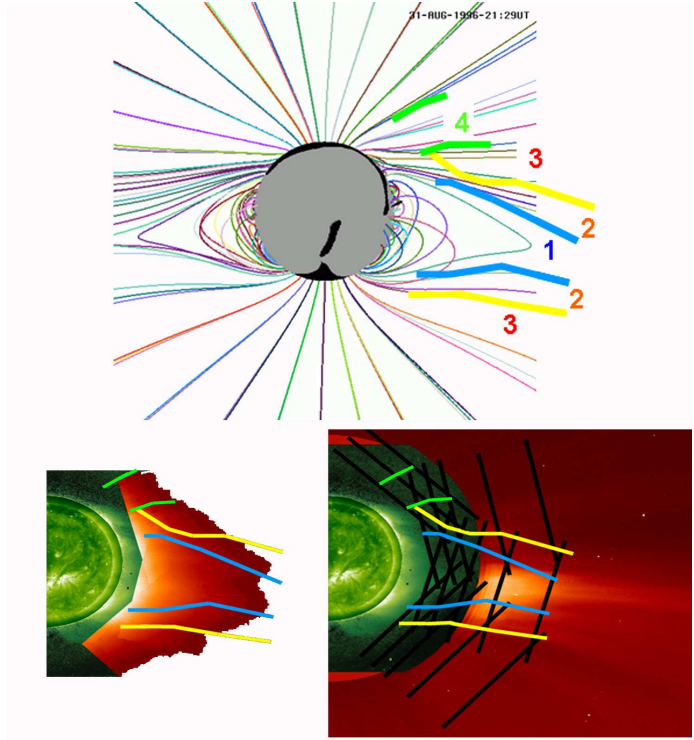}}
\caption{\small{
  {\bf{Top:}} Magnetic field lines map calculated by the 3D-MHD model with labeled regions and streamer boundaries defined
in the text.
The field lines shown are only those close to the plane of sky ($\pm$ 10 degrees). The coronal hole
boundary (black regions) calculated from the photospheric magnetic maps
is shown on the disk surface. {\bf{Bottom:}} Composite images of the Sun: the disk is imaged with EIT in the FeXII 195 \AA\, line (green, disk); the limb images are obtained with the LASCO C1 coronagraph (green, off-limb) and the coronal images are obtained with UVCS OVI 1032 line (interpolated red image) with superposed the streamer boundaries (left panel) and LASCO C2 visible light (red outer image) with superposed the slit of UVCS (black lines) and the streamer boundaries (right panel).
  \label{magn_map}}}
\end{figure}
In the case of the observation on August 31,
 the field of view of UVCS was
 well centered relative to the magnetic core of the streamer, derived from the MHD extrapolations.
 This is then selected for the analysis of the internal part of the
 streamer.
 The other dates of the UVCS
  observations, August 30 and September 1, 1996,
  are considered for the analysis of the open field regions just outside
  streamers. The observations on August 26 are used for the analysis of the regions at the edge of coronal hole.\\
 The streamer boundary is defined as coincident with the interface between closed and open
 field lines obtained from the extrapolations of the photospheric magnetic field for the dates of observation of
 the streamers. The magnetic streamer boundaries are shown in Figure 1 as blue lines. The precision of the determination of the boundary
 depends on the resolution used in the magnetic model to extrapolate coronal field lines and in this way it
 depends on the accuracy of the model itself.
 The boundaries are then compared with those defined on the basis of the intensity of UV lines in the outer corona
  (Abbo and Antonucci 2002 and Antonucci et al. 2005). The streamer boundaries defined 
as the 1/e 1032 O~VI peak intensity level, shown in Figure 1 as yellow lines, result to be external to those defined on the basis of the magnetic field model extrapolated from the photospheric data by $\sim$5-10$^o$. The 1/e intensity boundary has been considered in several previous work (e.g. Abbo et al. 2002, 2003a; Marocchi et al. 2001; Uzzo et al. 2003; Antonucci et al. 2005). This is a reasonable definition of the streamer boundary in absence of a direct inference of the magnetic topology of the coronal regions, as discussed in more details by Antonucci et al. 2005.
The bottom panels of Figure 1 show both the solar disk imaged with EIT (Fe XII 195 \AA) and the limb with LASCO C1 but, on the left, the coronal image is obtained with UVCS and,
on the right, by LASCO C2.
 In order to derive the plasma conditions within the streamer, and in the external regions,
the intensities of the O~VI 1032 and 1038 lines are integrated in four regions shown in Figure 1 on the top panel:
 within the streamer magnetic boundaries (region 1), in the intermediate region between the magnetic and the OVI intensity
  boundaries (region 2, as defined before), in the region external to the OVI intensity boundary (region 3, defined between $\sim$288$^o$ and $\sim$305$^o$) and in coronal hole (region 4, defined between $\sim$305$^o$ and $\sim$320$^o$).\\
  Stray light correction is applied and counts are transformed to intensity, $I(\lambda)$, by applying the
  standard radiometric calibration (Gardner et al. 1996).
  The integrated emissions are then fitted with a gaussian function, representing the coronal profile,
  convolved with a Lorentzian curve which accounts for the instrumental broadening and a rectangular
  function accounting for the width of the spectrometer slit.
  The function resulting from the convolution is added to a background linearly dependent on wavelength.
  The best fit is obtained by applying the least square method with the following quantities as
  adjustable parameters: standard deviation, $\sigma$, and mean wavelength, $\lambda_0$, of the solar profile,
  and slope and intercept of the background. Finally, the observed line intensity is derived as the integral over
  the gaussian line profile.\\
  The electron density and the outflow velocity are derived from the ratio of the
collisional
to radiative component of the oxygen O~VI 1032 line,
with the constraint of mass flux conservation, according to the method discussed in section \ref{diagn}.
The cross-section geometry of the flux tubes connecting the corona with the heliosphere, described by the expansion factor, is
derived from the extrapolations of the coronal magnetic fields of the MHD model.
Figure \ref{exp_fac} shows the expansion factors relative to the flux tubes of the streamer boundaries
 derived for the observations (dashed and dash-dotted lines show the North-West and the South-West 1/e boundary, respectively, and dash-dot-dotted and dotted lines show the North-West and the South-West mhd boundary, respectively),
  compared with the expansion factors derived from the model
  of Wang \& Sheeley (1990)(solid line),
  assumed in the analysis of Antonucci et al. (2005). We note that the increase around 2.5 \rsun~is less marked in the expansion factors
  derived by Miki\'c et al. (1999). It has been suggested that there is a strong empirical anticorrelation between wind
speed and expansion factors of magnetic flux tubes near the Sun (e.g. Levine et al. 1977; Wang and Sheeley 1990). Cranmer (2005) proposed a possible explanation of this effect by studying the critical points of the momentum function along flux tubes that range from the pole to the edge of the streamer belt. For what concerns the present work, the analyis of the latitudinal variation of the expansion factors will be investigated in details in a future work.\\
\begin{figure}[!htb]
\centerline{\includegraphics[height=.3\textheight]{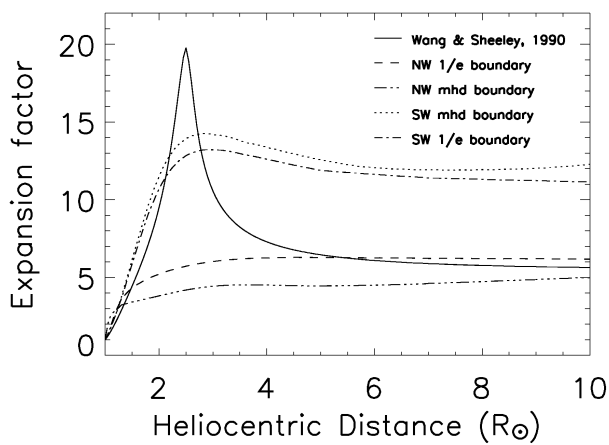}}
\caption{\small{Expansion factors relative to the flux tubes of the streamer boundaries as a function of
heliocentric distance: the solid curve refers to the model of Wang \& Sheeley (1990), the other curves refer
to model of Miki\'c et al. (1999) (dashed and dash-dotted lines show the North-West and the South-West 1/e boundary, respectively, and dash-dot-dotted and dotted lines show the North-West and the South-West mhd boundary, respectively), assumed in the present analysis.\label{exp_fac}}}
\end{figure}
\begin{figure}[!htb]
\centerline{\includegraphics[height=.3\textheight]{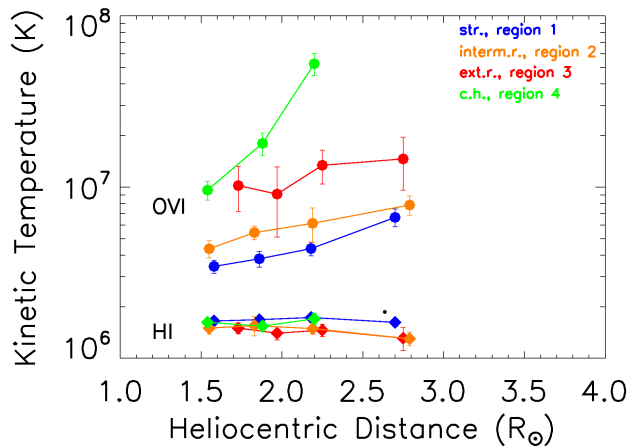}}
\caption{\small{Kinetic temperature of O$^{5+}$ ions and H$^0$ atoms as a function of heliocentric distance for streamer
 (region 1, blue points) for
region intermediate between the two boundaries (region 2, orange points), for the external region (region 3, red points) and for the base of coronal hole (region 4, green points).
\label{tk_th}}}
\end{figure}
The spectroscopic diagnostic technique can be applied when the distribution of the
oxygen ion velocity in three dimensions is determined, since, in a given volume, the absorption of the photons
 of the exciting spectrum is controlled by the coronal absorption profile along the incident direction in a solid
 angle subtending the disk of the Sun. The solar wind is assumed to be radial and the velocity
 distribution along the two directions perpendicular to the radial is considered to be the same.
The kinetic temperature of oxygen ions, $T_k$, expressed in terms of the spectral line width observed by UVCS,
 is a measure of the velocity distribution width
along the l.o.s.,
while for the
perpendicular directions to the l.o.s. has to be assumed (only for the fast wind in the core
 of coronal holes this quantity is partially constrained). Moreover, the derived kinetic temperature do not include the correction for removing the l.o.s. component of the outflow velocity.
 Figure 3 shows the kinetic temperature of O$^{5+}$ ions and H$^0$ atoms as a function of heliocentric distance for the four
 different regions as defined in Fig.1 (blue points for region 1, orange for region 2, red for region 3 and green for region 4).
 The data at 3.33 \rsun~are not included because, in this case, it is difficult to determine the true widths of the line profiles using 300 $\mu$m for the UVCS slit width.
It is reasonable to assume an isotropic
maxwellian velocity distribution inside the streamer, with the width defined by the observed $T_k$.
Within the streamer, the electron and ion densities are higher and due to the approximately static conditions
of the plasma, the isotropy of the ion velocity distribution is established via ion-ion collision.
Outside the streamer, in the regions 2, 3 and 4,
where there is an open field line topology, a bi-maxwellian velocity distribution
 of the ions is assumed, as for the analysis of coronal holes.
In this case, the ion kinetic temperature corresponds to the observed line width in the plane perpendicular
 to the
radial direction, and the ion kinetic temperature along the radial direction is equal to the electron temperature (maximum anisotropy).
 This assumption is dictated
by the fact that line broadening in the external regions (region 3) is about twice as larger
than inside the
streamer itself, thus representing an
 intermediate condition between closed field regions and the core region
of coronal holes, where the ion velocity distributions are found to be highly anisotropic
 (e.g. Kohl et~al., 1998,
Cranmer et~al., 1999b, Antonucci et~al., 2000a and b). Also in the regions between the boundaries (region 2) we
assume an anisotropic velocity distribution.\\
The coronal electron temperature, $T_e$, assumed in the analysis of streamers, is that
derived by
Gibson, et~al. (1999)
for the minimum of solar activity and varies between 1.5$\times 10^6$ K and 7$\times 10^5$ K in the
range of distance 1.5--3.5 R$_\odot$.
 We note that
the $T_e$ values do not influence significantly the results of the analysis
of electron density and outflow velocity (see also
 Antonucci et~al., 2000b). Therefore, in the absence of a direct measurement of $T_e$,
we assume in the open field line regions the same electron temperature of coronal holes,
 as inferred from the coronal hole measurements by David et~al. (1998) and extrapolated
to the values by Ko et al. (1997) further out in the corona obtained from $in-situ$ charge-state measurements performed with the SWICS experiment. That is, we assume
  that the regions outside streamers are cooler than the internal regions, with temperatures which never exceed 1$\times 10^6$ K.

\section{Results}

\begin{figure}[!htb]
\includegraphics[height=.4\textheight]{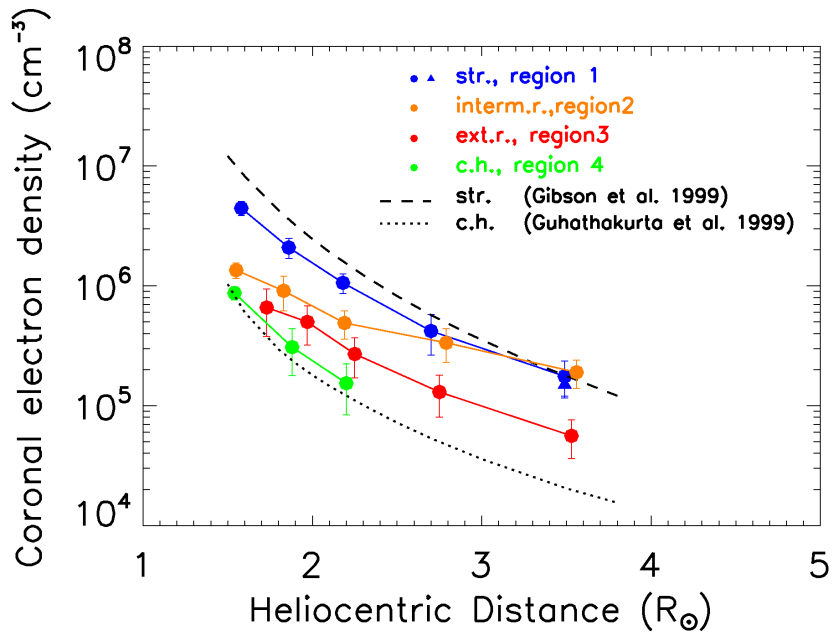}
\caption{\small{Electron density (cm$^{-3}$) as a function
of heliocentric distance (in solar radii), calculated
considering the
region within the streamer boundaries (blue dots, region 1), the region between the boundaries (orange dots, region 2), the
region external to the boundary defined as 1/e OVI intensity (red dots, region 3) and at the edge of coronal hole (green dots, region 4).
The dashed line shows the values derived by Gibson et~al. (1999) from visible light observations
 for streamers
 and the dotted line shows results obtained by Guhathakurta et~al. (1999) for coronal holes.
Static conditions are assumed in the streamer due to the presence of closed magnetic field; at 3.5 \rsun,
the blue triangle shows the result by assuming an expanding coronal plasma. The error bars have been estimated based on the propagation of the statistical uncertainties of the observed OVI 1032 and 1037 line intensities.
\label{ne_r}}}
 \end{figure}

\begin{figure}[!htb]
\includegraphics[height=.4\textheight]{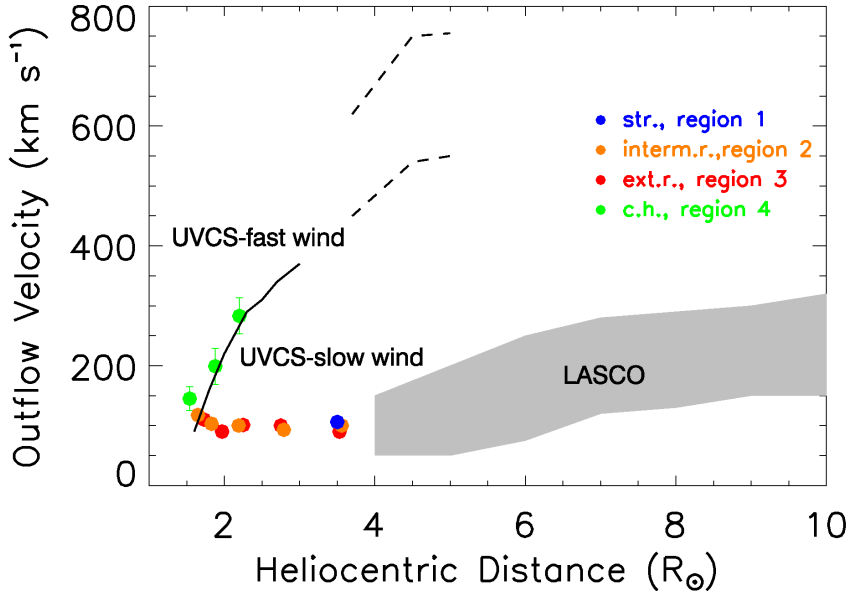}
\caption{\small{Outflow velocity (km/s) of the slow wind for the considered four regions as a function of the heliocentric distance (in solar radii).
The grey band from 4 to 10 R$_{\odot}$
shows the range of outflow velocities for the slow wind obtained with LASCO (Sheeley et~al., 1997). The solid curve up to 3 R$_{\odot}$~represents the
values of the fast wind obtained from the UVCS data (Antonucci et al., 2000b) and the dashed curves show the
 results by Telloni et al. (2007) of the fast wind velocity. The error bars (small in many cases) have been estimated based on the propagation of the statistical uncertainties of the observed OVI 1032 and 1037 line intensities.\label{vel_r}}}
 \end{figure}

We derive the electron density for the three
  regions inside and outside the streamer and for the region at the edge of the coronal hole as defined in Figure 1. The results are shown in Figure
   \ref{ne_r} and Table 2.
   The electron density values relative to the inner part of the streamers (region 1, blue dots)
   are calculated assuming a static plasma. The values are compatible with those derived in streamers
    from the visible light coronal observations by Gibson et~al. (1999)
     (dashed line). At 3.5 \rsun, we obtain also a value (blue triangle) by assuming an expanding coronal plasma
     and an anisoptropic velocity distribution.
 For what concerns the regions external to the streamer boundaries (region 3),
we have computed the electron density
taking into account the magnetic topology of the flux tubes as derived by the MHD model
 and the
 results are shown as red dots in Fig. 4,
compared with those derived for streamers by Gibson et~al.
(1999) (dashed line) and  for coronal holes by Guhathakurta et~al.
(1999) (dotted line). The electron density values relative to the region between boundaries (region 2) and at the base of the coronal hole (region 4)
 are shown, respectively, as orange and green dots. The error bars do not include uncertainties in the MHD model, but are based on  the propagation of the statistical uncertainties of the observed OVI 1032 and 1037 line intensities. The density and outflow velocity results however also depend on other factors, as discussed in Antonucci et al. 2004. The values in the regions 1 and 2 are comparable beyond 2.5 \rsun.
 This fact (see also the T$_{k,OVI}$ results in the previous section) suggests that the plasma of these regions could be the same, when the streamer converges to form
 the current sheet. The density results outside the streamer have intermediate values between those of a coronal hole and
 of the streamer
and they are compatible  with those obtained in the analysis of Antonucci et al. (2005). The region 4 is characterized by T$_{k,OVI}$ and  n$_e$ values of the coronal hole core.  Moreover, at 1.5
R$_{\odot}$ the electron density values in regions 2 and 3 are more similar to the coronal hole values than those in the streamer. These results are interesting for the analysis of the latitudinal variation of the physical parameters from slow to fast wind.\\
Figure \ref{vel_r} and Table \ref{tab:vout} show the values of the outflow velocities as a function of heliocentric distance for regions 2, 3 and 4 and the velocity at one height (3.5 R$_{\odot}$) for streamer region 1. In Figure \ref{vel_r},
   they are extended to higher altitudes by
  including the white-light
observations by LASCO (Sheeley et~al., 1997) in the range from 4 to 10 R$_{\odot}$\,(gray band). The values are much lower
 than those of the fast wind, which reaches 200 km/s at 2 R$_{\odot}$~and 400 km/s at 3 R$_{\odot}$.
     The solid curve
 represents the
values obtained from the UVCS analysis in the center of coronal holes up to 3 R$_{\odot}$~and the dashed curves show the
 results by Telloni et al. (2007) on the fast wind velocity with two different hypotheses for the degree of anisotropy in the velocity distribution.
  The outflow velocity values vary between 90 and 120 km/s approximately and there is no significant difference
  between the outflow velocity values in region 2 and in region 3, but there is a considerable change of O VI kinetic temperature and electron density values from one region to the other one. For what concerns region 1, within the streamer, we found at 3.5 \rsun~
  an outflow velocity value equal to 106$\pm$5 km/s. On the basis of the coronal field lines, at this height the current sheet is already formed. Hence, we find slow wind flowing along the axis of the streamer above the cusp. Finally, region 4 shows typical outflow velocity values of the coronal hole core. This fact indicates that, at the edge of coronal holes, the plasma has already reached the characteristics of the fast wind regime.\\

 \begin{table}[!htb]
   \begin{center}
    \caption{\small{Electron density averaged along the line-of-sight, $<n_e>$,
      derived for the three regions as a function of the height: the inner part of the streamer (region 1),
      the intermediate region between the two boundaries (region 2) and the region external to the 1/e
       intensity boundary (region 3).}}\vspace{1em}
     \renewcommand{\arraystretch}{1.2}
\begin{tabular}[h]{c}
\begin{tabular}[h]{|cc|cc|}
      \hline
      Region 1 & & Region 2 & \\
        r (R$_\odot$) &  $<n_e>$ (cm$^{-3}$) & r (R$_\odot$) &  $<n_e>$ (cm$^{-3}$)    \\
      \hline
     1.58&4.5$\pm$0.6 $\times$10$^6$ & 1.55&1.3$\pm$0.2 $\times$10$^6$ \\
     1.86& 2.1$\pm$0.4 $\times$10$^6$ & 1.83&9.1$\pm$2.9 $\times$10$^5$  \\
      2.18&1.1$\pm$0.2 $\times$10$^6$ & 2.19&4.9$\pm$1.3 $\times$10$^5$ \\
      2.70&4.2$\pm$1.5 $\times$10$^5$ & 2.79&3.3$\pm$1.0 $\times$10$^5$ \\
     3.49& 1.8$\pm$0.6 $\times$10$^5$ & 3.56&1.90$\pm$0.5 $\times$10$^5$ \\
    3.49$^1$ & 1.5$\pm$0.3 $\times$10$^5$ & & \\
      \hline
      Region 3 & & Region 4  &\\
        r (R$_\odot$) &  $<n_e>$ (cm$^{-3}$)  & r (R$_\odot$) &  $<n_e>$ (cm$^{-3}$)   \\
      \hline
      1.73&6.6$\pm$2.8 $\times$10$^5$ & 1.54 & 8.7$\pm$1.1 $\times$10$^5$\\
      1.97&5.0$\pm$1.8 $\times$10$^5$ & 1.88 & 3.1$\pm$1.3 $\times$10$^5$\\
      2.25&2.7$\pm$1.1 $\times$10$^5$ & 2.2 & 1.5$\pm$0.7 $\times$10$^5$\\
      2.75&1.3$\pm$0.5 $\times$10$^5$ & & \\
      3.53&5.6$\pm$2.0 $\times$10$^4$ & & \\
       \hline
      \end{tabular}
       \end{tabular}
     \label{tab:ne}
   \end{center}
 {\small{$^1$ with the assumption of an expanding coronal plasma.}}
 \end{table}
 \begin{table}[!htb]
   \begin{center}
    \caption{\small{Outflow velocity (km/s) of the slow wind as a function
     of heliocentric distance (in solar radii) for the intermediate and the external regions
      (see text for details).}}\vspace{1em}
    \renewcommand{\arraystretch}{1.2}
     \begin{tabular}[h]{|cc|cc|cc|}
      \hline
      Region 2 && Region 3 && Region 4&\\
        r (R$_\odot$) & v$_{out}$ (km/s) & r (R$_\odot$) & v$_{out}$ (km/s)& r (R$_\odot$) & v$_{out}$ (km/s)\\
      \hline
      1.65& 117.5$\pm$10&1.73& 110$\pm$5&1.54&145$\pm$20\\
     1.83& 103$\pm$10&1.97& 90$\pm$10&1.88&199$\pm$30 \\
      2.19& 100$\pm$5&2.25&101$\pm$5&2.2&283$\pm$30\\
      2.79&93$\pm$5&2.75&100$\pm$5&&\\
     3.56& 100$\pm$5&3.53&90$\pm$10&&\\
      \hline
       \end{tabular}
    \label{tab:vout}
   \end{center}
 \end{table}

 \section{Conclusions}
The results of kinetic temperature, electron density and ion outflow velocity in the
streamer belt of the minimum corona (1996) lead to a more accurate identification of the sources of the
  slow coronal wind than in previous analyses. The slow wind is found, mainly, to flow adjacent to
the streamer boundary in the open magnetic field line region. Moreover, the presence of outflowing plasma
  has been detected where the heliospheric current sheet is forming, approximately beyond 3.5 R$_\odot$. 
   These results while confirming, to a large degree, those of Antonucci et al. (2005), show that the region where the slow wind is observed is more extended toward the streamer, since the mhd boundary is closer to the streamer axis than the 1/e instensity boundary. In the previous paper,
the authors suggested that the open field lines running along the streamer boundary originate in the same
 unipolar magnetic region lying at the base of a polar coronal hole, because the regions just outside the streamer boundary are characterized by the same density as the hole edge below 2 R$_\odot$. The present analysis confirms this hypothesis:
  the density values of regions 2,3 and 4 are comparable around 1.5 \rsun, while further out in the corona, the density varies
   significantly from the coronal hole edge toward the streamer.\\
 The outflow velocity values are comparable with those derived by Sheeley et~al. (1997),
 from the white-light observations of LASCO between 2 R$_{\odot}$\,and 5 R$_{\odot}$. The values of electron density and
 outflow velocity in the streamer adjacent regions are consistent with those found by Antonucci et al. (2005) up to
 2.75 \rsun. At 3.5 \rsun~the values of $<n_e>$ are lower in the present analysis with consequently higher
  outflow velocity values. We note that in this range of heights (2.5--3.5 \rsun), the coronal expansion factors assumed
  in the two studies have a different behaviour (see Fig. 2).
  The outflow velocity results of the present analysis
 are consistent with those found by Strachan et al. (2002):
 they analysed the visible and
  ultraviolet coronal emission of an equatorial streamer, using the Doppler dimming technique,
  along the streamer axis from 2.3 R$_{\odot}$ to 5.1 R$_{\odot}$, and
  across the streamer, but only at 2.33 R$_{\odot}$.
  They obtained outflows as a function of height along the streamer
axis with values ranging from about 50 to 100 km/s from 4.1 to 5.1 solar
radii. There were no detectable outflows along the streamer axis below
3.5 R$_{\odot}$. Outflow
velocities from 50-110 km/s were detected only outside $\sim$15 degrees from
the streamer axis at 2.33 R$_{\odot}$. In our analysis of the regions along the streamer axis,
   the outflow velocities are higher, in the range within (100-110 km/s),
   and are observed at
  lower heights, at 3.5 R$_{\odot}$. The difference
  between the results of the two studies can be traced to the fact that
  the values of the inferred electron densities are different. In this study n$_e$ and w are derived using the same observations. The density used by
  Strachan et~al. (2002) is inferred from visible light data which usually yields higher
  values.\\
  The results of our analysis suggest that the slow wind flows externally to the bright parts of the streamer
  called 'legs', proposed as coronal sources of the slow wind by i.e. Raymond et al. (1997) and Uzzo et al. (2003)
  and there is no detectable wind contribution coming from these structures, thus it can be deduced that they are characterized by a closed magnetic topology. The scenario resulting from the identification of the slow
 wind sources with the regions surrounding the streamer boundaries and possibly with the
 region close to the current sheet, is  compatible  with the model proposed by
 Wang et~al. (2000), of a  two--component slow wind: one component
 flowing  along the rapidly diverging open magnetic field lines adjacent to
 the streamer boundary, and the second one confined to the region of the
 denser equatorial plasma sheet. The latter case has the same observational
 characteristics as the model of Noci et~al. (1997).

\section{Acknowledgments}
UVCS is a joint project of the
National Aeronautics and Space Administration (NASA), the Agenzia
Spaziale Italiana (ASI) and Swiss Founding Agencies.
The research of LA has been funded through the contract
 I/023/09/0 between the National Institute for Astrophysics (INAF)
 and the Italian Space Agency (ASI) and also through the contract ASI/I/035/05/0,
by Agenzia Spaziale Italiana and Ministero dell'Istruzione, dell'Universit\`a
e della Ricerca. The work at Predictive Science, Inc. was partially supported by NASA's
Heliophysics Theory and SR \& T programs, and NSF's CISM program.

\end{document}